# Intrusion Detection: A Deep Learning Approach

Ishaan Shivhare [a], Joy Purohit [b], Vinay Jogani [c], Samina Attari [d] and Dr. Madhav Chandane [e]

[a,b,c,d,e] Department of CE & IT, Veermata Jijabai Technological Institute, Mumbai, India.

ishaanshivhare2001@gmail.com, purohitjoy21@gmail.com, joganivinay@gmail.com, sameenaattari7860@gmail.com and mmchandane@it.vjti.ac.in

*Abstract*—Network intrusions are a significant problem in all industries today. A critical part of the solution is being able to effectively detect intrusions. With recent advances in artificial intelligence, current research has begun adopting deep learning approaches for intrusion detection. Current approaches for multi-class intrusion detection include the use of a deep neural network. However, it fails to take into account spatial relationships between the data objects and long term dependencies present in the dataset. The paper proposes a novel architecture to combat intrusion detection that has a Convolutional Neural Network (CNN) module, along with a Long Short Term Memory(LSTM) module and with a Support Vector Machine (SVM) classification function. The analysis is followed by a comparison of both conventional machine learning techniques and deep learning methodologies, which highlights areas that could be further explored.

*Keywords*—Intrusion Detection, Long Short Term Memory (LSTM), Convolutional Neural Network (CNN), Support Vector Machine (SVM).

## I. INTRODUCTION

In the years recently passed, due to the fast emergence and development of information and communication technologies, network security has become an increasingly important aspect of everyone's daily lives. As the internet becomes increasingly ubiquitous, widespread technologies are contributing to the massive data scale of the world of today. The cost of cybercrime to the world economy is over $1 Trillion and is only expected to grow. Moreover, traditional anomaly-based detection methods can no longer be applied to today's Internet. Thus researchers have begun adopting more advanced techniques in the cybersecurity domain.

Effective intrusion detection is a very important component of a strong anti-intrusion system. Many scientists have studied the application of artificial intelligence to develop a highly accurate intrusion detection system. The methods that are generating the most interest today generally involve machine learning or deep learning-based approaches. However, one of the shortcomings of a traditional machine learning-based approach is that these systems mainly analyze features of network traffic that have been extracted manually. Deep learning-based systems can analyze these same features and also extract features from the original traffic automatically, thus circumventing this problem and enhancing the accuracy of the intrusion detection system.

Some researchers have explored the use of artificial neural networks to perform intrusion detection Vinayakumar et.al [1]. This approach is effective in detecting unforeseen and unpredictable intrusive attacks and is a flexible system. More recently Laghrissi et. al [2] discusses the use of a variant of a recurrent neural network for the IDS. This is due to the advantages realized by the extraction of temporal features by the Long Short Term Memory (LSTM) model.

In order to further enhance the performance of the LSTM-based model, Sun et.al [3] demonstrate the effectiveness of using a convolutional neural network (CNN) in conjunction with an LSTM, to leverage the extraction and analysis of spatial features as well as temporal features of the network traffic data. This paper proposes a modified version of the previously mentioned work for the task of intrusion detection. The model will use a CNN-LSTM model but introduce a linear support vector machine (SVM) that replaces the final output layer of the Softmax function.

## II. LITERATURE REVIEW

Jha et.al [4] uses a Support Vector Machine based system for detecting intrusions. Using a novel approach the best features for intrusion detection are selected, the accuracy of the model is increased. This approach combines filter and wrapper models for the selection of relevant features. The training time is also reduced. Ahmad et.al [5] conducts a comparative analysis of different classical machine-learning models on the NSL-KDD dataset. The results depict that SVM is a good choice of algorithm for the task of intrusion detection.

Vinayakumar et.al [1] demonstrate the advantage of a deep learning-based approach to intrusion detection compared to old machine learning classifiers. Due to the vulnerable nature of malware, the attack methods are constantly changing. Deep learning-based approaches are more flexible and can classify unforeseen and unpredictable cyberattacks more effectively. Moreover, such systems can circumvent the problem of extracting features manually from the network traffic which is a limitation of classical machine learning system-based approaches. The paper compares the accuracy of Deep Neural Networks (DNN) of different depths to other standard machine learning approaches on a variety of datasets, thus confirming that the DNN learns and works well with the abstract and high-dimensional feature representations of IDS data.

Laghrissi et.al [6] uses a Long Short-Term Memory (LSTM) based model to detect intrusions. LSTM is a deep learning-based approach and is a subset of the Recurrent Neural Network (RNN). The use of Principal Component Analysis (PCA) as a technique for dimensionality reduction is also demonstrated in the paper, and the combination of PCA and LSTM obtained the best results on the KDD99 dataset as compared to when other dimensionality reduction methods such as Mutual Information (MI) were used.

Agarap et.al [7] propose an intrusion detection system that is an amendment to a variation of a DNN-based approach. The system consists of a variation of the LSTM called a Gated Recurrent Unit (GRU). A GRU is less complex than an LSTM and is useful in cases where a small dataset is to be utilized. Conventionally, both of the previously mentioned RNN variants use the Softmax function for the final output layer in order to make the prediction and the cross-entropy function to calculate the loss. The proposed system replaces the Softmax function with a linear SVM and the cross-entropy with a margin-based function. The proposed system outperforms a

GRU-Softmax model on the task of binary classification on intrusion detection using the Kyoto dataset. SVM is better than Softmax in performance, training, and testing time due to low computational complexity.

The paper Sun et.al [3] proposes a deep learning-based intrusion detection system that is a combination of Convolutional Neural Network (CNN) and LSTM. The paper proposes that combining the two would improve the extraction of spatial and temporal elements from the data, improving intrusion detection systems even further. The accuracy of the CNN-LSTM model is compared with a CNN-only model, an LSTM-only model, and other machine learning models on the CICIDS 2017 dataset, where the CNN-LSTM model achieves the best performance out of all the others.

III. PROPOSED SYSTEM

A. Flowchart

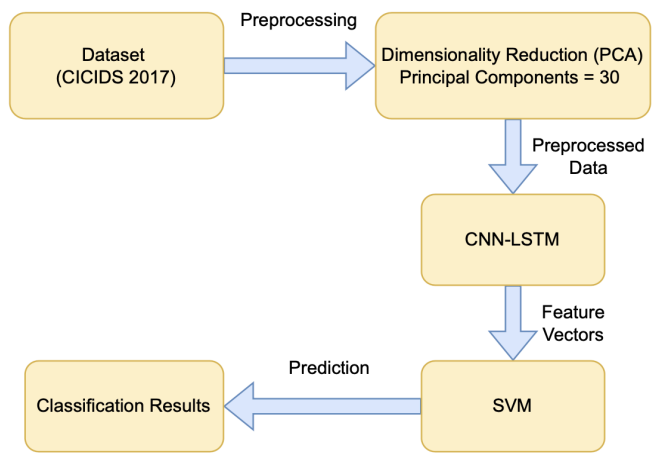

Figure 1: Flowchart of the proposed model

B. Problem Statement

Based on the research conducted in the literature review, a model for the task of intrusion detection is proposed. The relevant flowchart diagram can be observed in figure 1.
The modules contained in the proposed model are as follows:

1. CNN for extraction of spatial features.
2. LSTM for extraction of temporal features.
3. SVM to perform the final classification in place of the traditional softmax output layer.

In addition to these modules the exploration and implementation of dimensionality reduction of the data using Principal Component Analysis (PCA) to further enhance the performance of the model is conducted.

C. Architecture

The proposed system leverages the use of a CNN-LSTM-SVM based approach. This can be observed in the architecture diagram provided in figure 2.

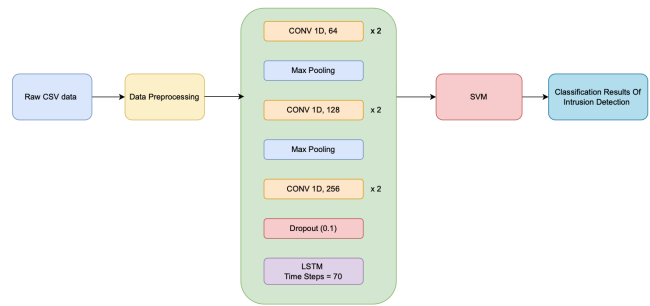

Figure 2: Architecture Diagram of the proposed model

1. CNN

The CNN module in the proposed system consists of 3 layers of convolution blocks, where max pooling is conducted between each block. Following this, a dropout layer is integrated which takes inputs from the last convolution block in order to avoid overfitting. These feature vectors extracted after the dropout layer are then fed to an LSTM module.

The usage of a CNN aids the model in learning the features and underlying concepts more efficiently, by extracting spatial features that contribute to a better overall performance of the intrusion detection system.

2. LSTM

The LSTM module receives feature vectors from the dropout layer and consists of 70 time steps to learn the interdependencies in the features.

Integrating LSTM in the deep learning architecture enables efficient extraction of temporal data elements. It is able to thus learn the context of the hidden features leading to intrusions. Hence, it aids the model in classifying such temporal intrusions effectively.

3. SVM

The feature vectors extracted from the CNN-LSTM model are then fed to a SVM classifier consisting of a Radial Basis Function (RBF).

SVM as a traditional machine learning algorithm demonstrated high performance for the task of multi-class classification of network traffic data. This coupled with the fact that the usage of SVM in place of a traditional Softmax function layer would decrease computation time, increase efficiency and scalability make it a strong candidate for the final output layer.

IV. IMPLEMENTATION DETAILS

A. Dataset

The CICIDS2017 [9] has been used. This dataset includes both benign ("normal") traffic and anomalous ("attacks") traffic. This dataset contains 14 different forms of attacks. The most recent and benign common attacks are included in the CICIDS2017 dataset, which closely mirrors actual real-world data. Brute Force FTP, Brute Force SSH, DoS, Heartbleed, Web Attack, Infiltration, Botnet, and DDoS are some of the attacks performed. Thus, the main reason for choosing the CICIDS-2017 dataset was motivated by the need to have a dataset that accurately depicts the actual network activity today in the experiments.

B. Data Preprocessing

1. Feature subset selection and dropping null and strange values.

'Destination port' feature has been dropped as it does not provide any useful information to the prediction of the model. Furthermore, all the data entries having null and strange values (for eg. inf,-inf) have been cleaned.

2. Feature normalization

Min-max normalization is performed on all values of the attributes in the dataset to avoid providing unfair weightage to any feature element.

3. Dimensionality reduction

After feature subset selection, the data still had 77 attributes which would immensely add to the computational complexity of the model. Hence, Principal Component Analysis (PCA) [10] was performed to reduce the dimensionality of the data.

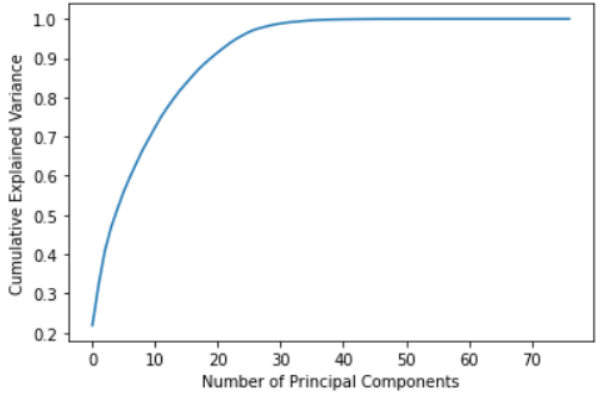

*Figure 3. Selecting the optimum number of principal components*

As illustrated in the figure above, the maximum variance in the data was captured by a minimum of 30 principal components. Hence, 30 principal components are enough to explain most of the variance in the data. So using these results, the dataset of 77 features is transformed to a dataset having 30 features using PCA which reduces the computational complexity significantly.

V. RESULTS AND ANALYSIS

Table 1. Comparative analysis of the performance on the test set CICIDS 2017 on the task of Multi-Class classification

| Models | Accuracy | Increase in accuracy (%) |
|---|---|---|
| KNN (k=5) | 90.1% | 7.19 |
| Random Forest | 88.48% | 8.81 |
| CNN | 91.65% | 5.64 |
| CNN-LSTM | 93.61% | 3.68 |
| DNN (5 Layers) | 95.61% | 1.68 |
| CNN-LSTM-SVM (proposed model) | 97.29% | - |

Table 1 shows a comparison of the effectiveness of deep learning methods (CNN, CNN-LSTM, and CNN-LSTM-SVM) and traditional machine learning methods (KNN, RF) on the test dataset. Furthermore, an increase in classification accuracy when using the proposed model compared to other models can be observed. It can be seen that the proposed method (CNN-LSTM-SVM) obtains a better accuracy on the testing set. Hence, this proves that the proposed method has achieved superior performance as compared to other techniques, validating the advantages provided by the approach described in Section 3.3 above.

VI. CONCLUSION

To summarize, in this paper first a detailed literature review of the current research is conducted on intrusion detection systems using deep learning. Then a model is proposed on the basis of this research that leverages the spatial and temporal feature extraction ability of CNN-LSTM and the decrease in training time and the increase in performance and scalability of an SVM classifier to give a higher-performing intrusion detection system. The CICIDS 2017 dataset is used to test and train the model, on the task of multi-class classification where the model proved to be prolific.

VII. FUTURE SCOPE

In this paper, the CNN-LSTM-SVM model has been proposed to improve the network intrusion detection system's accuracy. However, only one kind of LSTM has been tested in this work. Future research can examine many LSTM variations, including Peephole, Multiplicative, and Weighted LSTM, as well as other neural network and feature selection techniques.